\documentclass[PRL,preprint,amsmath,amssymb,superscriptaddress,longbibliography]{revtex4-1}

\usepackage{graphicx}% Include figure files untitled folder
\usepackage{dcolumn}% Align table columns on decimal point
\usepackage{bm}% bold math
\usepackage{hyperref}
\usepackage{breakurl}
\usepackage{multirow}
\usepackage{enumitem}
\begin{document}

\title{Defects controlled hole doping and multi-valley transport in SnSe single crystals}

%and multi-valley transport

\author{Zhen Wang}
       \thanks{Equal contributions}
      \affiliation{Department of Physics, Zhejiang University, Hangzhou 310027, P. R. China}
       \affiliation{State Key Lab of Silicon Materials, Zhejiang University, Hangzhou 310027, P. R. China}

\author{Congcong Fan}
      \thanks{Equal contributions}
      \affiliation{State Key Laboratory of Functional Materials for Informatics, Shanghai Institute of Microsystem and Information Technology (SIMIT), Chinese Academy of Sciences, Shanghai 200050, China}
      \affiliation{CAS Center for Excellence in Superconducting Electronics (CENSE), Shanghai 200050, China}

\author{Zhixuan Shen}
      \thanks{Equal contributions}
      \affiliation{Department of Physics, Zhejiang University, Hangzhou 310027, P. R. China}

\author{Chenqiang Hua}
      \affiliation{Department of Physics, Zhejiang University, Hangzhou 310027, P. R. China}

\author{Qifeng Hu}
      \affiliation{Department of Physics, Zhejiang University, Hangzhou 310027, P. R. China}

\author{Feng Sheng}
       \affiliation{Department of Physics, Zhejiang University, Hangzhou 310027, P. R. China}

\author{Yunhao Lu}
      \affiliation{State Key Lab of Silicon Materials, Zhejiang University, Hangzhou 310027, P. R. China}

\author{Hanyan Fang}
      \affiliation{Department of Chemistry, National University of Singapore, 3 Science Drive 3, Singapore 117543}

\author{Zhizhan Qiu}
      \affiliation{Department of Chemistry, National University of Singapore, 3 Science Drive 3, Singapore 117543}

\author{Jiong Lu}
      \affiliation{Department of Chemistry, National University of Singapore, 3 Science Drive 3, Singapore 117543}

\author{Zhu-An Xu}
      \affiliation{Department of Physics, Zhejiang University, Hangzhou 310027, P. R. China}
      \affiliation{State Key Lab of Silicon Materials, Zhejiang University, Hangzhou 310027, P. R. China}
      \affiliation{Zhejiang California International NanoSystems Institute, Zhejiang University, Hangzhou 310058, P. R. China}
      \affiliation{Collaborative Innovation Centre of Advanced Microstructures, Nanjing 210093, P. R. China}

\author{D. W. Shen}
       \email{dwshen@mail.sim.ac.cn}
      \affiliation{State Key Laboratory of Functional Materials for Informatics, Shanghai Institute of Microsystem and Information Technology (SIMIT), Chinese Academy of Sciences, Shanghai 200050, China}
      \affiliation{CAS Center for Excellence in Superconducting Electronics (CENSE), Shanghai 200050, China}

\author{Yi Zheng}
      \email{phyzhengyi@zju.edu.cn}
      \affiliation{Department of Physics, Zhejiang University, Hangzhou 310027, P. R. China}
      \affiliation{Zhejiang California International NanoSystems Institute, Zhejiang University, Hangzhou 310058, P. R. China}
      \affiliation{Collaborative Innovation Centre of Advanced Microstructures, Nanjing 210093, P. R. China}

\date{\today}

\begin{abstract}

SnSe is a promising thermoelectric material with record-breaking figure of merit, \textit{i.e., ZT}. As a semiconductor, optimal electrical dosage is the key challenge to maximize \textit{ZT} in SnSe. However, to date a comprehensive understanding of the electronic structure and most critically, the self-hole doping mechanism in SnSe is still absent. Here, we report the highly anisotropic electronic structure of SnSe investigated by both angle-resolved photoemission spectroscopy and quantum transport, in which a unique ``\textit{pudding-mold}'' shaped valence band with quasi-linear energy dispersion is revealed. We prove that the electrical doping in SnSe is extrinsically controlled by the formation of SnSe$_{2}$ micro-domains induced by local phase segregation. Using different growth methods and conditions, we have achieved wide tuning of hole doping in SnSe, ranging from intrinsic semiconducting behaviour to typical metal with carrier density of $1.23\times 10^{18}$ cm$^{-3}$ at room temperature. The resulting multi-valley transport in $p$-SnSe is characterized by non-saturating weak localization along the armchair axis, due to strong intervalley scattering enhanced by in-plane ferroelectric dipole field of the puckering lattice. Strikingly, quantum oscillations of magnetoresistance reveal three-dimensional electronic structure with unusual interlayer coupling strength in $p$-SnSe, which is correlated to the interweaving of SnSe individual layers by unique point dislocation defects. Our results suggest that defect engineering may provide versatile routes in improving the thermoelectric performance of the SnSe family.
\end{abstract}

\maketitle

Although the syntheses and electrical measurements of the Group-IV monochalcogenides SnS, SnSe, GeS and GeSe can be dated back to 1956 \cite{Mooser_PR1956_SnSe}, the interests in this family soar very recently after the report of an extraordinary thermoelectric conversion efficiency of $\textit{ZT}=2.6$ at 923 K in SnSe \cite{ZhaoLD_14SnSe_Nature}. The dimensionless value of $\textit{ZT}$ is known to be complex interplay between the Seebeck coefficient ($S_{xx}$), the electrical conductivity ($\sigma_{xx}$), and the thermal conductivity ($\kappa$), as defined by the formula of $\textit{ZT}=(S_{xx}^{2}\sigma_{xx}/\kappa)T$. The unprecedented $\textit{ZT}$ in SnSe has been attributed to an ultralow $\kappa$, due to a giant phonon anharmonic effect \cite{ZhaoLD_14SnSe_Nature}, which is supported by inelastic neutron scattering measurements \cite{Delaire_15SnSeAnharmonic_NatPhy}. Noticeably, there is a second-order phase transition in SnSe at  $\sim$810 K, from the low-temperature $Pnma$ (\#62) phase to the high-temperature $Cmcm$ (\#63) phase \cite{Schnering_86SnSePhase_JPCS}. The $Cmcm$ phase has a small direct bandgap of 0.46 eV, which is not sufficient to suppress thermal activation of electrons in the conduction band. With strong bipolar transport above 700 K, $\kappa$ in SnSe is dominated by the electronic contribution ($\kappa_{el}$) rather than the anharmonic phonon part ($\kappa_{L}$) \cite{Kioupakis_15SnSeZT_JAP}. It is thus proposed by Shi and Kioupakis to introduce hole doping to maximize the thermoelectric performance in SnSe, which has an upper limit of $\textit{ZT}=S_{xx}^{2}\sigma_{xx}T/\kappa_{el}$ \cite{Kioupakis_15SnSeZT_JAP}. Using sodium as hole dopants, Zhao \textit{et al.} have observed drastic \textit{ZT} increase in $p$-SnSe \cite{ZhaoLD_16SnSe_Science}. Compared to other doped thermoelectric materials \cite{Smirnov70_SemiLeadChalcogen,Snyder_11PbSeSxx_AdvMat}, $S_{xx}$ in Na-doped SnSe is unusually high \cite{ZhaoLD_16SnSe_Science}, which is a strong indication of multiple pocket thermal transport. Yet, the physical original of such multi-valley transport is not clear, since theoretical calculations just suggest single-band dominated $S_{xx}$ at the experimental doping level. Electron doping of SnSe by iodine \cite{RenZF_15SnSeIodine_AdvEnerMater} and bismuth \cite{ChoS_16SnSeBi_NatComm} is less impressive with a highest \textit{ZT} value of 2.2 \cite{ChoS_16SnSeBi_NatComm}, although theoretical models predict unrivalled thermoelectric performance in $n$-type SnSe \cite{HuangBL_15SnSenDoping_PRB,Kuroki_17SnSeDFT_arXiv}.

\begin{figure*}[!thb]
\begin{center}
\includegraphics[width=6.5in]{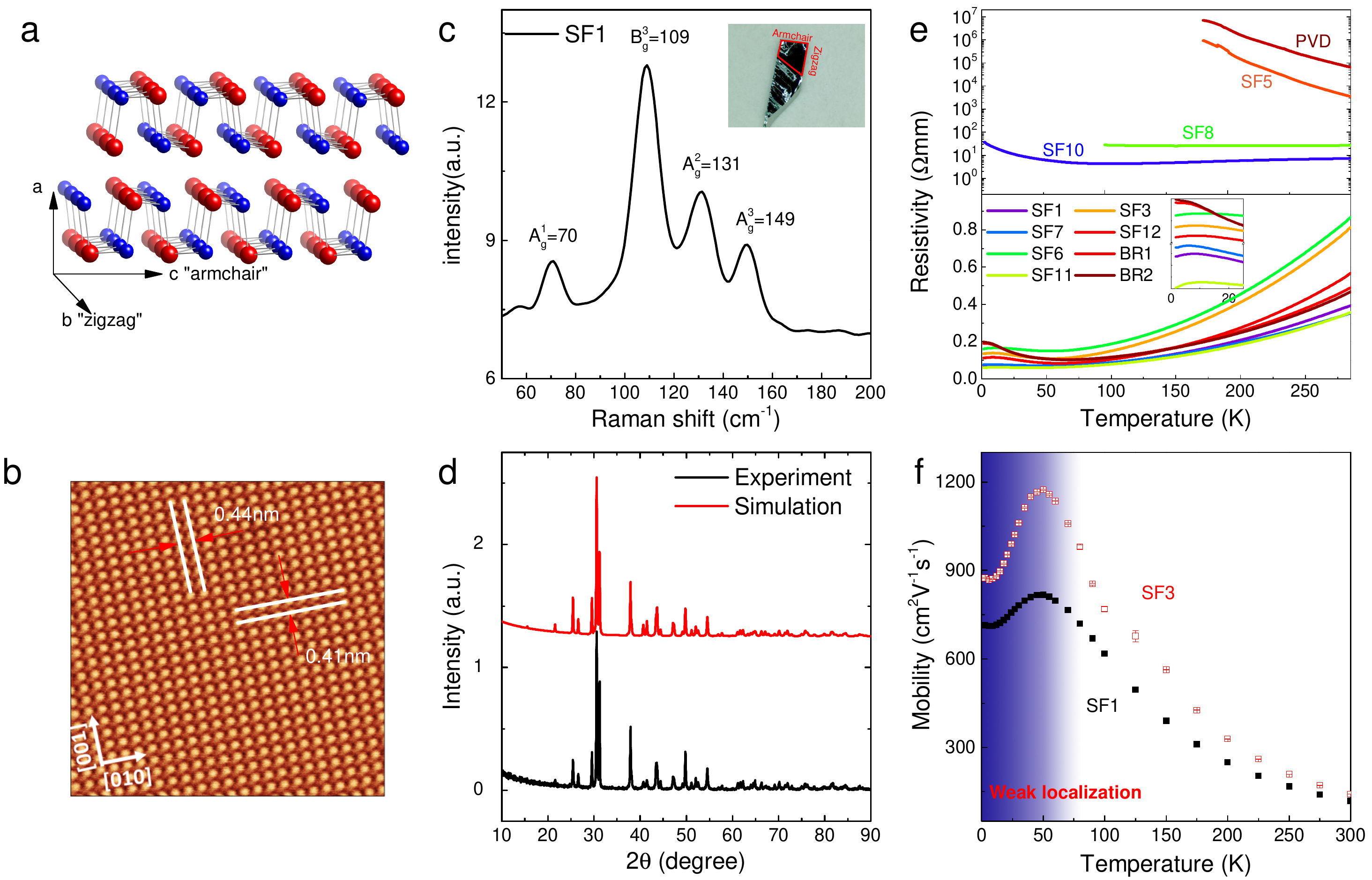}
\end{center}
\caption{\textbf{\label{Fig1} Characteristics of SnSe single crystals.} (\textbf{a}) SnSe crystallizes in the layered structure as a binary analogue of BP, by forming alternating Sn- and Se-rows along the zigzag axis.  The red and blue balls represents Sn and Se atoms respectively. (\textbf{b}) High-resolution nc-AFM image of stoichiometric SF1 SnSe taken at 4.3 K. (\textbf{c}) RT Raman spectrum of SnSe shows fingerprinting phonon modes. The inset shows a typical sliced single crystal of SnSe, whose crystallographic axes are determined by XRD. Note that the large natural edge of the sample is along the diagonal of the armchair and zigzag axes. (\textbf{d}) XRD patterns of ground polycrystal of SF1 SnSe, agreeing well with the simulation. (\textbf{e}) $T$-dependent $\rho$ curves for all batches of crystals, showing various conduction behaviour from metal (lower panel) to semiconductor (upper panel). Below 50 K, all metallic batches exhibit upturns in $\rho$ as temperature decreases, a manifestation of the quantum interference effect, $i.e.$, weak localization (WL). The inset zooms in the resistivity curves below 30K. (\textbf{f}) $T$-dependent Hall measurements reveal metallic behaviour for SF1 (black) and SF3 SnSe (red). The mobility drops below 50 K confirms WL in both batches of samples, consistent with the $\rho-T$ results.}
\end{figure*}

Despite of the critical importance of electrical doping, the low-lying electronic structure of SnSe, which is mainly composed of the resonantly bonded $p$-orbitals and intimately correlated to the formation of strongly anisotropic lattice anharmonicity \cite{Delaire_15SnSeAnharmonic_NatPhy}, is not well studied by experiments. Most critically, the as-synthesized SnSe is moderately hole doped, instead of semiconducting with an predicted indirect band gap. A fundamental understanding of the self-hole doping mechanism may allow the community to develop an effective doping strategy for SnSe, and reconcile the discrepancy between theoretical predictions and experiments.

In this work, we report the low energy electronic structure of SnSe single crystals, and elucidate the physical origins of the hole doping mechanism and striking three-dimensionality of the Fermi surface in two unique types of defects, namely local phase segregation and point dislocations, respectively. Using high-resolution angle-resolved photoemission spectroscopy (ARPES), for the first time we have found that the low-lying valence band of the as-grown $p$-SnSe exhibits the unique ``\textit{pudding-mold}'' shape with quasi-linear dispersion, which could compellingly account for the high $S_{xx}$ in this material. We show unambiguous evidences that locally segregated SnSe$_{2}$ micro-domains play crucial roles in determining the observed hole doping in SnSe. By utilising different growth methods and fine tuning growth parameters, we demonstrate that the extrinsic hole doping in SnSe can be widely tuned from semiconducting to $1.23\times 10^{18}$ cm$^{-3}$. The ARPES deduced non-parabolic band parameters and multi-valley Fermi surface are well supported by the transport results of Shubnikov-de Haas oscillations in magnetoresistance (MR), which shows distinct non-saturating weak localization along the armchair axis. Such highly anisotropic quantum localization phenomena can be enhanced by increasing hole doping, suggesting the importance of strong intervalley scattering assisted by in-plane ferroelectric dipole field of the puckering lattice. Strikingly, quantum oscillations also reveal unusual interlayer coupling strength and three-dimensional Fermi surface in $p$-SnSe, contradicting to the two dimensional nature of the BP-type layered structure. Using atomic force microscopy, it is found that SnSe monolayers are heavily interwoven by high intensity of point dislocations, which retain the antiferroelectric stacking order in the bulk but interconnect two second-nearest neighbouring monolayers with the same dipole orientation by point defects.

% \rowcolor{myblue}
%  \multicolumn{9}{|c|}{\multirow{-1}{*}{ \textbf{TABLE I: Physical properties of SnSe single crystals grown by three different methods}}}\\

\section{Results}
\textbf{Single crystal growth and characterization.} Figure \ref{Fig1}a illustrates the layered structure of SnSe, which is essentially a binary analogue of black phosphorus \cite{ZhangYB_14BPFET_NatNano} with alternating Sn (red) and Se (blue) rows forming puckering bilayer structure in the $b-c$ plane. Two different crystallographic directions of armchair ($c$-axis) and zigzag ($b$-axis) can be clearly distinguished. As summarized in Table \ref{table:growth}, we have synthesized 12 batches of SnSe single crystals using three growth methods, including self flux (SF), Bridgeman (BR) and physical vapour deposition (PVD), combined with various growth parameters [see Supplementary Informations (SI) for the details of single crystal growth for different batches]. For stoichiometrically synthesised SnSe, vacancies are rarely seen under $q$-plus based non-contact atomic force microscopy (Figure \ref{Fig1}b, details in Methods and SI). Noticeably, SnSe crystals tend to form large edges along the diagonal axis of the $b-c$ plane. It is thus crucial to determine the crystallographic orientations of individual samples by X-ray diffraction (XRD) before physical characterizations and transport measurements (see SI). The inset of Figure \ref{Fig1}c shows a typical sliced SnSe single crystal, with shining surface and physical dimensions of $a=1\,\mathrm{mm}$, $b=6\,\mathrm{mm}$, and $c=3\,\mathrm{mm}$. Raman spectroscopy shows four fingerprinting phonon modes of $A_{g}^{1}$, $B_{g}^{3}$, $A_{g}^{2}$, and $A_{g}^3$ for the $Pnma$ phase (Figure \ref{Fig1}c), which is consistent with previous reports \cite{Cardona_77SnSenRaman_PRB,LiuZF_15SnSeCVD_NanoRes}. The high quality of our single crystals is also confirmed by the XRD results of the ground polycrystal sample, in which the characteristic Bragg diffraction peaks perfectly match the \#62 space group (Figure \ref{Fig1}d).

Remarkably, without the introduction of any external dopants, different batches of SnSe single crystals exhibit distinct conduction behaviours, ranging from semiconductivity to metallicity. In a short summary, the flux cooling rates from the highest growth temperature ($950 \,^{\circ}{\rm C}$) determine the hole doping levels and thus the resistivity of SnSe. For SF and BR growth, the fast cooling in 24 hours to $400 \,^{\circ}{\rm C}$ (SF1, SF6, SF7, SF11, SF12, BR1, and BR2) leads to metallic behaviour, while the slow cooling in 7 days to $400 \,^{\circ}{\rm C}$ (SF8) produces semiconducting SnSe with thermal activation behaviour. Using optical microscopy and atomic force microscopy, we have identified two unique types of defects in SnSe, namely point dislocations and SnSe$_{2}$ micro domains embedded in SnSe single crystals, respectively (see SI). Unexpectedly, there is a strong correlation between the conduction behaviour in SnSe and the concentrations of SnSe$_{2}$ micro crystals, which are typically several micrometers in lateral sizes and several MLs to several tens MLs in thickness (see SI). With a threshold of $\sim$0.3\% SnSe$_{2}$, the resistivity vs temperature ($\rho-T$) curves of SnSe becomes metallic over the whole $T$ range from 300 K to 1.5 K (the lower panel of Figure \ref{Fig1}e), while semiconducting SnSe shows negligible concentration of SnSe$_{2}$ micro domains (the upper panel of Figure \ref{Fig1}e). The formation of SnSe$_{2}$ micro domains can be explained by local phase segregation in Se-rich areas. To test the hypothesis, we have synthesized two extra batches of SnSe crystals, by grinding stoichiometric Sn-Se powder in argon atmosphere for 1 hour before the growth. As shown in Table \ref{table:growth}, the grinding step effectively suppresses inhomogeneity in the growth flux, and no concentration of SnSe$_{2}$ is observed in the slow cooling batch of SF5. For the fast-cooling crystals (SF10), very low ratio of SnSe$_{2}$ can still be found, which is not surprising since SnSe$_{2}$ nucleations ($\sim 650\,^{\circ}{\rm C}$) may start during the late stage of crystal growth. We have further verified the claim by the PVD method, in which SF1-type single crystals containing $\sim$1\% SnSe$_{2}$ are ground into polycrystal powder and used as the PVD source. In a temperature gradient of $800-700\,^{\circ}{\rm C}$ over 12 cm, a pure phase of SnSe has been obtained by the sublimation and re-deposition process, while SnSe$_{2}$ was removed due to the low melting point.

\begin{table*}
\caption{Physical properties of SnSe single crystals grown by three different methods}
\begin{center}
  \begin{tabular}{ |p{2cm}<{\centering}  p{2.5cm}<{\centering} p{1.8cm}<{\centering} p{0.8cm}<{\centering} p{1.8cm}<{\centering}  p{1.8cm}<{\centering} p{0.8cm}<{\centering} p{1.8cm}<{\centering}  p{1.8cm}<{\centering} |}
  \hline
   \textbf{Batches} & \textbf{Flux cooling rate}& \textbf{SnSe$_{2}$  ratio$^{\ast}$} & \textbf{Se:Sn}&\multicolumn{2}{c}{\textbf{n} $(\textbf{10}^{\textbf{18}} \textbf{cm}^{\textbf{-3}})$}& & \multicolumn{2}{c|}{\textbf{Resistivity} ${(\boldsymbol{\Omega} \textbf{mm}})$} \\
  \cline{5-6}
  \cline{8-9}
   &  &   &  & $\textbf{SdH}$ $^{\dag}$ & $\textbf{Hall}^{\textbf{RT}}$ & & $\textbf{2K}$ &$\textbf{275K}$\\
  \hline
  SF7 &Fast cooling& $3.73\%$ &1& $0.836$ & $1.23$ & &  $0.0776$ & $0.333$\\
  SF1 &Fast cooling& $0.9\%$ &1& $0.825$ & $1.16$ & &  $0.0746$ & $0.360$\\
  SF6 &Fast cooling& $0.3\%$ &1& $0.704$ & $0.69$ & &  $0.162$ & $0.812$\\
  SF3 &3 days to 673K& $<0.3\%$ &1& $0.502$ & $0.592$ & &  $0.136$ & $0.757$\\
  SF11 &Fast cooling& $2.6\%$ &1& $0.480$ & $1.079$ & &  $0.0604$ & $0.333$\\
  SF12 &Fast cooling& $<0.3\%$ &0.95& $0.472$ & $0.794$ & &  $0.191$ & $0.457$\\
  BR1 &Fast cooling& $2.49\%$ &1& $0.423$ & $0.699$ & &  $0.113$ & $0.525$\\
  BR2 &Fast cooling& $\sim0.5\%$ &1& $0.395$ & $0.591$ & &  $0.198$ & $0.436$\\
  SF8 &7 days to 673K& $<0.1\%$ &1& -- &  -- & &  -- & $26.4$\\
  SF10$^{\ddag}$ &Fast cooling& $<0.1\%$ &1&-- &  -- & &  $36.74$ & $6.93$\\
  SF5$^{\ddag}$ &3 days to 673K& $None$ &1& -- &  -- & &  -- & $8303$\\
  PVD &--& $None$ &1&-- &  -- & & -- & $137760$\\
  \hline
  \end{tabular}
 \end{center}
  \label{table:growth}
  $^{\ast}$Coverage of SnSe$_{2}$ after the microexforliation of surface layers. Statistics by 500x optical microscope.

  $^{\dag}$$n_{SdH}$ is calculated by $\frac{2k_{SdH}^3}{3\pi^2}$, assuming spherical FSs. A factor of two is multiplied to reflect the double valley transport.

  $^{\ddag}$ Sn and Se are ground and thoroughly mixed before the growth.
\end{table*}

%(\textbf{a (c)}), and (\textbf{b (d)}), , in comparison with SF3 (\textbf{a}) and (\textbf{b})
We have noticed that SnSe single crystals with high $ZT$ values reported in the literatures all belong to the metallic type in our study. In particularly, the RT hole doping of the SF3 batch ($\sim 6\times10^{17}cm^{-3}$) matches the reference report \cite{ZhaoLD_14SnSe_Nature}, which holds the highest reported $ZT$ value. The experimental $S_{xx}$ of SF3 SnSe ($\sim 570\,\mu V/K$ at 300 K. See SI) also shows striking consistency with the crystals used in Ref. \cite{ZhaoLD_14SnSe_Nature}. By cooling down the SF3 samples to 1.5 K, we have observed that the $\rho-T$ curve shows a minimum at around 50 K, followed by an upturn down to the base temperature. By measuring the Hall signals, we determined that the $\rho-T$ behaviour from RT to 50 K is mainly due to hole mobility enhancement at low temperatures (Figure \ref{Fig1}f), which is typical for metals, but not consistent with thermally activated charge transport in semiconductors. Such metallic behaviour is general for different batches of hole-doped SnSe with significant presence of SnSe$_{2}$ micro domains. As a comparison, we have also systematically studied the $T$-dependent charge transport of the SF1 batch, which have a factor of two higher hole doping of $1.23\times10^{18}\,\mathrm{cm}^{-3}$. As shown in Figure \ref{Fig1}f, hole mobility of the SF1 sample shows the same $T$-dependence as SF3, despite that the highest mobility point at 50 K is 1.5 times larger in the latter. The resistivity upturn of hole-doped SnSe below 50 K is a transport signature of weak localization (WL) \cite{WLreview_Lin_JPCM02}, as we will clarify later in the quantum transport part.

\begin{figure*}[!thb]
\begin{center}
\includegraphics[width=6.5in]{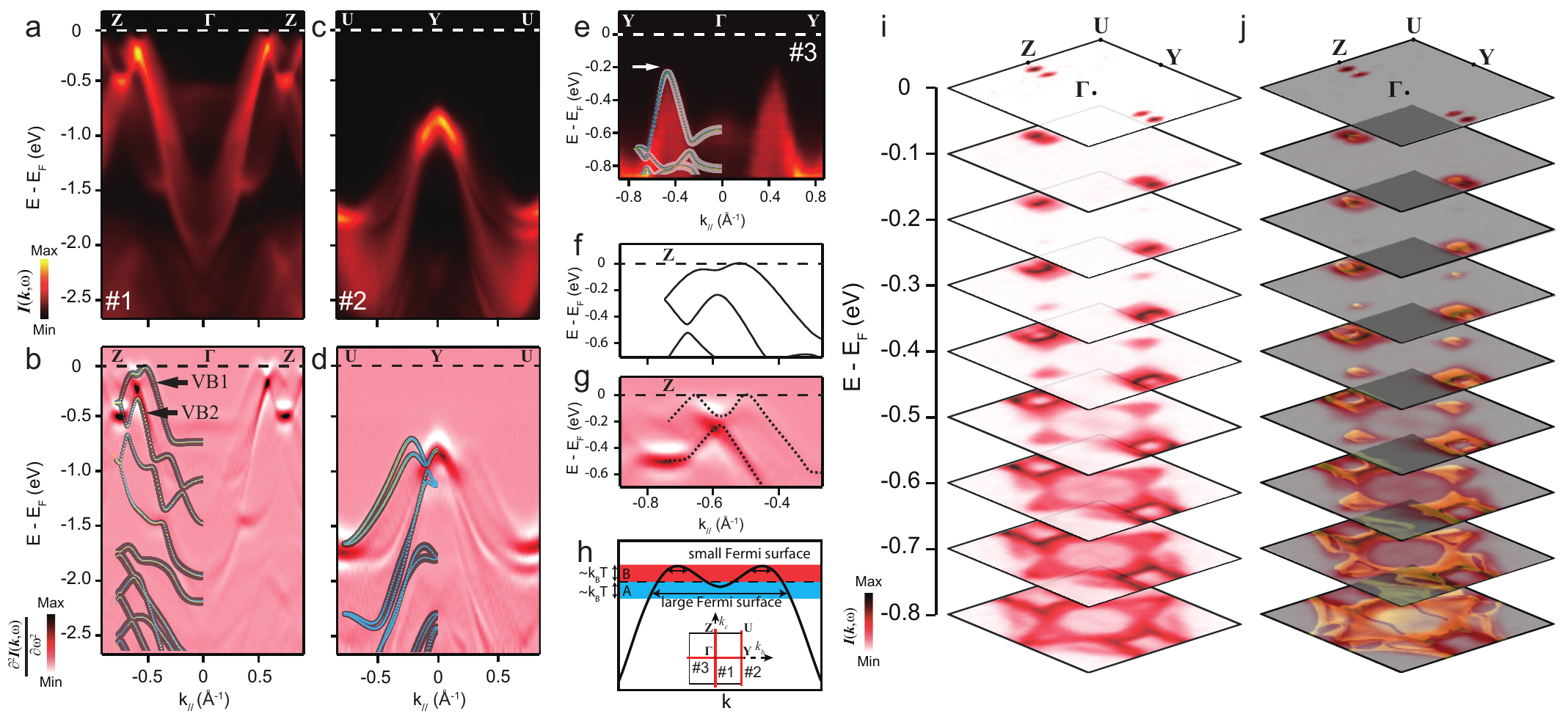}
\end{center}
\caption{\textbf{\label{Fig2} ARPES resolved electronic structure of $p$-SnSe, compared with DFT calculations.} ARPES measured band dispersions along high-symmetry directions $Z-\Gamma-Z$ (\textbf{a}), $U-Y-U$ (\textbf{c}), and $Y-\Gamma-Y$ (\textbf{e}), taken with 50 eV photons. The second derivative plots (\textbf{b}) and (\textbf{d}), corresponding to (\textbf{a}) and (\textbf{c}) respectively, are compared with the DFT results, in which the weight of Sn 5s- and Se 4p-orbitals are represented by light blue and yellow colours, respectively.  A close-up of the top VBs along $\Gamma-Z$ reveals pronounced differences between the theoretical band dispersion (\textbf{f}) and the ARPES measured dispersions (\textbf{g}), as highlighted as the dashed lines. (\textbf{h}) A schematic plot of a pudding-mold shaped band with corrugation, which leads to giant $S_{xx}$ due to the band geometry effect. The inset of (\textbf{h}) indicates the ARPES cut directions in the projected two-dimensional BZ. (\textbf{i}) and (\textbf{j}) Comparison of stacked plots of constant-energy contours at different $E_{B}$, which show good agreement between the experiments and calculations below 0.2 eV. Black squares here represent the boundary of the first BZ.}
\end{figure*}

\textbf{Electronic structure of $p$-SnSe.} To understand the metallic behaviour in $p$-SnSe single crystals and the corresponding correlation with their high $ZT$ values, we performed systematic ARPES measurements to investigate their band structure (see Methods for details). Figures \ref{Fig2}a-\ref{Fig2}e summarize the direct comparison of experimental band dispersions and DFT calculations along high-symmetry directions. Overall, as a weakly correlated system, the ARPES results of SnSe valence bands (VBs) can be well reproduced by DFT after a rigid shift of chemical potential, indicating that randomly distributed SnSe$_{2}$ micro domains just play the role of charge reservoir but do not affect the electronic structure of SnSe much. This argument could be further confirmed by the close resemblance between the experimental and DFT constant energy contour plots (Figures \ref{Fig2}i and \ref{Fig2}j). As shown in Figure \ref{Fig2}a and \ref{Fig2}b, ARPES measurements reveal two small contiguous hole-like pockets along $\Gamma-Z$ in the vicinity of the Fermi level ($E_F$), forming two side-by-side ellipsoid Fermi pockets (Figure \ref{Fig2}i), which is in good agreement with the metallic transport properties of these samples. By projecting the Sn and Se orbitals to the DFT-calculated VBs, we find that this highest VB (VB1) is dominated by Se 4$p$-states, with a relatively low weight of the Sn 5$s$ orbital (Figure \ref{Fig2}b). However, the detailed band structure in the vicinity of the $E_F$ of SnSe still shows some pronounced differences between the DFT and ARPES results. For DFT, as highlighted in Figure \ref{Fig2}f, the band top of VB1 is found along $\Gamma-Z$ with the Brillouin-zone positions of (0.00, 0.00, 0.52). Moving toward the zone boundary, there is another local VB maximum located at (0.00, 0.00, 0.64), which is yet about 50 meV lower than the band top. However, the ARPES determined VB maxima of this band are nearly equal in energy (Figure \ref{Fig2}g), leading to the formation of the unique pudding-mold shaped band.

As illustrated in Figure \ref{Fig2}h, a pudding-mold band is characterized by a relatively flat portion at the top which bends sharply to the quickly dispersive lower portion. Consequently, when the chemical potential is moderately tuned to be around the bending point, $S_{xx}$ would become unusually large due to a rapid change in charge carrier mobility across $E_F$ ~\cite{Arita_07PuddingBand}. In the case of SF1 SnSe presented here, such band geometry effect could be more drastically enhanced by the quasi-linear band dispersions of VB1, as highlighted by the X-shaped dashed lines in Figure \ref{Fig2}g. By plotting constant energy contours, we can naively evaluate the FS evolution as function of rigid chemical potential shifts ($E_{B}$), which is equivalent to doping changes, and thus get insights into the thermoelectric performance of heavily $p$-doped SnSe. As shown in Figure \ref{Fig2}i and \ref{Fig2}j, multiple sets of VBs could be activated in charge transport by gradually increasing hole doping. These include another local maximum of VB1 along $\Gamma-Y$ and the VB2 valley for $E_{B}\geqslant 0.2$ eV, and a giant flat band at the zone center when $E_{B}$ reaches $\sim 0.6$ eV. For the highest experimental hole doping of $4\times 10^{19}\,\mathrm{cm}^{-3}$ \cite{ZhaoLD_16SnSe_Science}, we can estimate a $E_{B}$ of 0.13 eV by assuming a constant effective mass of 0.2 $m_{e}$, which is still distant from the saddle point (0.18 eV) of the pudding-mold VB. It may thus conclude that the observed usually high $S_{xx}$ for Na-doped SnSe is a combination of band geometry effect and multi-valley thermal transport.

\begin{figure*}[!thb]
\begin{center}
\includegraphics[width=5.5in]{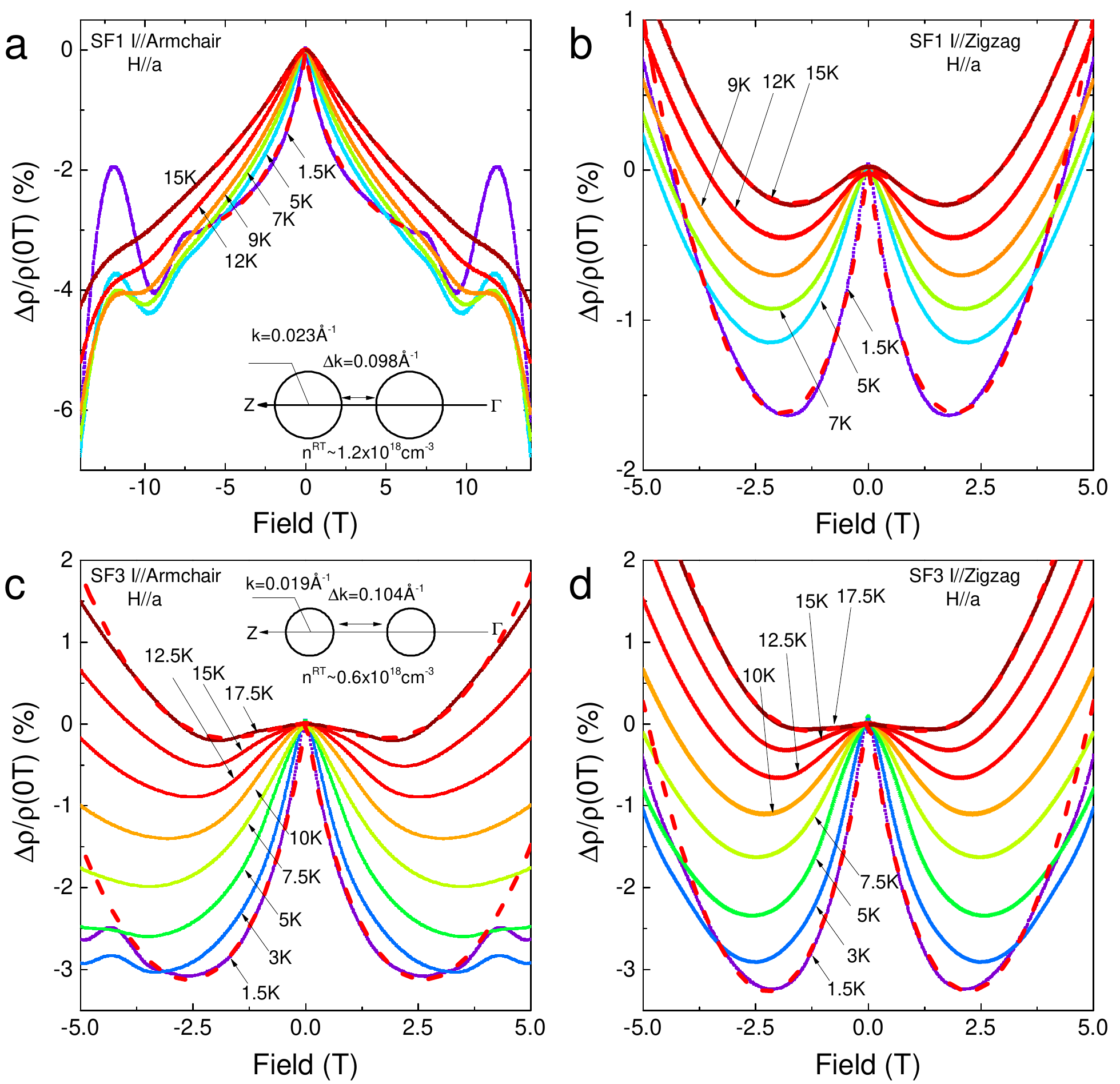}
 \end{center}
\caption{\textbf{\label{Fig3} Doping-dependent anisotropic weak localization in $p$-SnSe.} (\textbf{a}) and (\textbf{b}), WL induced NMR in SF1 SnSe for $I\parallel b$ and $I\parallel c$, respectively. The red-dash lines are the fitting results using the WL model (see Eqn. \ref{eq1}), which explains the evolution of NMR very well by $T$-dependent $l_{\phi}$. Strikingly, the NMR of SF1 SnSe for $I\parallel c$ is non-saturating up to 14 T, while NMR for $I\parallel b$ dwindles as a function of $T$ and is only dominant at low fields below 2 T. Such exotic WL behaviour show strong doping dependence, as evident by weaker NMR in SF3 SnSe [(\textbf{a}) and (\textbf{b})]. The insets of (\textbf{a}) and (\textbf{c}) show the sketches of constant-energy contours of FSs for SF1 and SF3 respectively. Intervalley scattering assisted by the in-plane dipole field along $\Gamma-Z$ is enhanced when the two pudding-mold valleys become closer.}
\end{figure*}

\textbf{Quantum localization in $p$-SnSe.} For $p$-SnSe in this study, the doping levels are well below $2\times 10^{18}$ cm$^{-3}$, corresponding to a chemical potential which is located closely below the VB1 maxima. Thus, the charge transport in our samples are also dominated by the multi-valley FSs of the pudding-mold shaped VB1. Such multi-valley transport in hole-doped SnSe is characterized by highly anisotropic WL, which is sensitive to hole doping levels. As shown in Figure \ref{Fig3}a, when the current flow ($I$) is aligned with the $c$ axis, the WL-induced negative magnetoresistance (NMR) for $B\parallel a$ in SF1 SnSe is not saturating, even in the Landau quantization regime above 5 T at 1.5 K. The non-saturating NMR, which persists up to the maximum field of 14 T, becomes more distinguishable at elevated temperature above 10 K, when SdH oscillations are effectively suppressed (Figure \ref{Fig3}d). Such exotic WL behaviour for $I\parallel c$ and $B\parallel a$ are in stark contrast to the MR characteristics for $I\parallel b$ and $B\parallel a$, in which NMR magnitudes dwindle as a function of $T$ and are only dominant at low fields below 2 T before normal positive MR prevails (Figure \ref{Fig3}b). Such an anisotropic WL behaviour can be readily reproduced in different batches of SnSe samples, but with a strong doping dependence. As shown in Figure \ref{Fig3}e, with the same configurations of $I\parallel c$ and $B\parallel a$, SF3 SnSe shows significantly weaker WL, which reaches a maximum $\Delta \rho/\rho$ change of about $-3\%$ at 2.5 T. By slightly warming up the sample to 5 K, parabolic positive MR becomes significant above 3 T (Figure \ref{Fig3}c). For this sample, WL for $I\parallel b$ and $B\parallel a$ is very similar to SF1 SnSe.

The doping-dependent anisotropic WL phenomena in $p$-SnSe are in coincident with a strong in-plane dipole field along the $\Gamma-Z$ direction, which originates from the binary BP lattice \cite{ZengXC_16SnSe_NanoLett, QianXF_17SnSeMultiferro_2DMater}. For multi-valley FSs in $p$-SnSe, such a strong dipole field may cause strong intervalley scattering to change the charge transport characteristics. Using the ARPES results, we have compared the FSs of SF1- and SF3-SnSe. As show in the insets of Figure \ref{Fig3}a and Figure \ref{Fig3}c, with 2 times higher hole doping in SF1 SnSe, the Fermi energy is  by $\sim5$ meV, which reduces the separation between the two pudding-mold valleys from $0.104\,A ^{-1}$ to $0.098\,A ^{-1}$ \footnote{Although the average reduction in the intervalley separation is not very impressive, the phase segregation may lead to non-uniform doping in SnSe, making the two pudding-mold valleys even closer in localized areas.}. Consequently, intervalley scattering is expected to be enhanced, considering that each SnSe monolayer is ferroelectric with a net dipole filed along the armchair direction \cite{ZengXC_16SnSe_NanoLett, QianXF_17SnSeMultiferro_2DMater}. For non-relativistic fermions, an increase in intervalley scattering generally leads to the suppression of WL, which can be quantitatively modelled by a simplified model\cite{kawabata1980theory} of
\begin{equation*}\label{eq1}
\frac{\Delta\rho}{\rho^2}=-\alpha\frac{e^2}{2\pi^2\hbar}\sqrt{\frac{eB}{\hbar}}F(\frac{4eBl_{\phi}^2}{\hbar})+cB^2,
\end{equation*}
in which $\alpha \leqslant 1$ is a fitting coefficient representing the overall WL effect, $B$ is the magnetic field, $l_{\phi}$ is the phase coherence length, $F(x)$ is the Hurwitz zeta function (See SI), and $cB^{2}$ is the normal parabolic MR. Indeed, by fitting the $T$-dependent WL in SF1- and SF3-SnSe samples, we consistently get $\alpha_c<\alpha_b$ and $l_{\phi-c}<l_{\phi-b}$, suggesting the suppression of WL by the highly anisotropic intervalley scattering between the two pudding-mold valleys. It is noteworthy that at 1.5 K, the $cB^{2}$ term is extremely weak along the armchair direction. This could also be related to the anisotropic intervalley scattering in SnSe, when the conventional parabolic MR due to the cyclotron movement of charge carriers became suppressed by the dipole field.

\begin{figure*}[!thb]
\begin{center}
\includegraphics[width=6.5in]{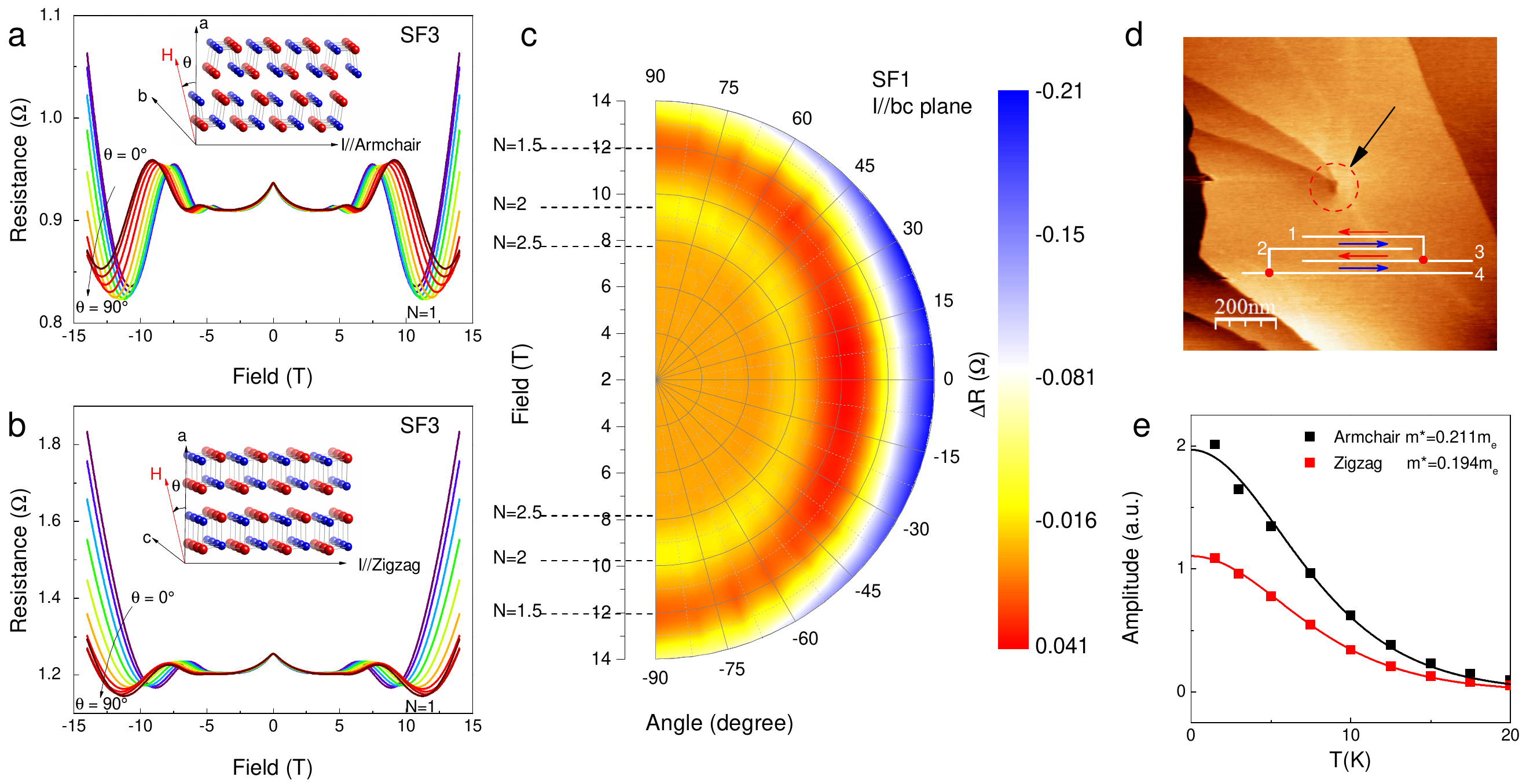}
\end{center}
\caption{\textbf{\label{Fig4} 3D Fermi surface in $p$-SnSe induced by point dislocations.} (\textbf{a}) and (\textbf{b}) Angle-dependent SdH oscillations of SF3 SnSe at 1.5 K for $I\parallel c$ and $I\parallel b$, respectively. The insets explain the measurement configuration, in which transverse field is rotated from the $a$ axis to the in-plane direction. (\textbf{c}) Polar contour plot of the $\theta$-dependent SdH oscillations in SF1 SnSe. The corresponding 3D FS is a elongated ellipsoid in the $b-c$ plane. (\textbf{d}) Typical nc-AFM image of point dislocations in SnSe crystals, showing the interconnection between the top ML and the third ML, while the stacking sequence between the top and second ML is reversed crossing the point defect. (\textbf{e}) LK-model fit of the $T$-dependent SdH amplitudes for SF3 SnSe, yielding similar effective mass around $0.2\,m_e$ for both directions.}
\end{figure*}

\textbf{Three dimensional Fermi surface of $p$-SnSe.} We further determined the FS morphology of SF1 and SF3 SnSe using angle-dependent SdH oscillations by rotating the samples in transverse $B$ with two current geometries of $I\parallel c$ and $I\parallel b$. Unexpectedly, we have resolved ellipsoid FSs with moderate anisotropy for both batches of samples, as evident by the angular evolution of the SdH peaks at 1.5 K. By fixing $B\parallel a$ ($\theta=0$), SdH oscillations mainly compare the anisotropy in charge carrier mobility along the $c$ and $b$ axes, respectively. As shown in Figure \ref{Fig4}a and \ref{Fig4}b, SdH oscillations in SF3 SnSe quickly reach the quantum limit (Landau level $N=1$) at 8.8 T for $I\parallel b$, while 10.8 T is required for $I\parallel c$. The observed anisotropic quantization for $I\parallel c$ and $I\parallel b$ is rooted in the puckering lattice of SnSe, which has higher mobility for the zigzag axis than the armchair direction for $p$-doped crystals \cite{Kioupakis_15SnSeZT_JAP}. To our surprise, the angle ($\theta$) dependence of SdH oscillations for both current configurations are rather weaker when the transverse field is rotated from the out-of-plane ($B\parallel a$) to the in-plane direction ($B\parallel b$ or $B\parallel c$). In Figure \ref{Fig3}c, we plotted the polar contour of the $\theta$-dependent SdH oscillations for SF1 SnSe with $I\parallel bc$, which also reveals a 3D FS independent of current flow. Such experimental results in SnSe can be directly compared to the two-dimensional quantum oscillations in BP \cite{ZhangYB_14BPFET_NatNano,ZhangYB_15BPSdH_NatNano,ZhangYB_16BPQHE_NatNano}, which shares the same puckering bilayer structure of resonantly bonded $p$-orbitals. The absence of a $1/cos(\theta)$ dependence in the SdH oscillation frequencies (see SI for detailed analysis) and the robust quantum oscillations for in-plane $B$ strongly suggest that self-doped $p$-SnSe has considerable interlayer coupling to suppress the two-dimensional nature of the BP-type lattice, which is theoretically predicted to have marginally higher interlayer coupling energy than graphite \cite{QianXF_17SnSeMultiferro_2DMater}.

Although SnSe$_{2}$ microcrystals play a critical role in determining the resistivity of SnSe, it is unlikely that the unusually enhanced interlayer coupling in SnSe is correlated to the local phase segregation. This is straightforward to see since two-dimensional SnSe$_{2}$ micro domains are randomly intercalating between the puckering SnSe monolayers, which destroys the stacking order of SnSe crystal. Instead, we found that point dislocations may be responsible for the strong interlayer coupling and 3D FS in SnSe. As shown in Figure \ref{Fig4}d, the formation of point dislocations is due to the interconnection between SnSe MLs with the same even/odd sequence. This type of unique interlayer connection by a point defect is the direct consequence of antiferroelectric stacking between SnSe MLs, which is thermodynamically favoured by the bulk phase (see SI). As schemed in the inset of Figure \ref{Fig4}d, within the $Pnma$ phase, neighbouring SnSe MLs have opposite in-plane ferroelectric dipole alignment, which prohibit the interconnection of even-odd MLs without destroying the antiferroelectric stacking order. On the other hand, point defects connecting the second-nearest neighbouring MLs with the same dipole orientation retain the antiferroelectric bulk phase by reversing the stacking sequence in the vicinity of a point dislocation (see SI for detailed illustrations). For SF1 SnSe, the dislocation density is larger than $4.5\times10^{4}\,\mathrm{mm}^{-2}$, which corresponds to more than one dislocation site in a randomly searched area of $10\times10\,\mathrm{\mu m}^{2}$ (see SI for AFM statistics). Such a high density of interlayer defects interweave SnSe MLs along the $a$ axis, and thus greatly weaken the two-dimensionality of SnSe. The three-dimensionality of FS induced by point dislocations is also manifested in effective mass ($m^{\ast}$), which is a quantitative parameter representing the band dispersion along different crystallographic axes. By fitting the $T$-dependent SdH amplitudes using the Lifshitz-Kosevich (LK) formula, we can get the effective mass for different $I$ and $B$ configurations (Figure \ref{Fig4}e). As summarized in Supplementary Table S1, $m^{\ast}$ deduced by quantum oscillations are comparable between the interlayer direction and the in-plane axes. Such band parameters are well supported by the ARPES results, which yield an effective mass of $m^{\ast}=0.258 m_{e}$ for the ``pudding-mold'' VB.

\section{Discussion}
We have used synchrotron radiation based ARPES and quantum transport measurements to probe the electronic structure of SnSe. The first-time resolved ``pudding-mold'' VB with quasi-linear energy dispersion may hold the key to understand the extraordinary thermoelectric performance of $p$-SnSe. Within the experimental hole doping levels achieved to date, charge transport in $p$-SnSe is contributed by two pudding-mold pockets, which could explain the multi-valley nature of the observed unusual thermoelectric transport in Na-doped SnSe. Equally important, the ultra-high $S_{xx}$ in $p$-SnSe also benefits from the band geometry of the ``pudding-mold'' VB, which causes drastic change in charge carrier mobility across $E_{F}$. Interestingly, by further increasing hole doping to $E_{B}=0.18$ eV, we would reach the saddle point of the pudding-mold VB. It is also tempting to explore $S_{xx}$ in $p$-SnSe at higher $E_{B}$ of 0.2 eV and 0.6 eV, where two sets of extra valleys will be activated in thermoelectric transport. However, this would require micro-mechanical exfoliation and device fabrications so that thickness-dependent study of SnSe becomes feasible. By introducing heavy ionic-liquid doping in SnSe thin-film devices, the ultimate thermoelectric performance of SnSe could be systematically explored. In the two-dimensional limit, such SnSe electronic devices would also be fascinating platforms to explore exotic physical phenomena, such as quantum Hall effects \cite{ZhangYB_16BPQHE_NatNano} and multi-ferroelectric properties \cite{ZengXC_16SnSe_NanoLett, QianXF_17SnSeMultiferro_2DMater}.

The origin of the pudding-mold VB, which significantly deviates from the DFT calculations, is beyond the scope of this study. Nevertheless, it may be due to the hybridisation of the Se $p$-orbitals and Sn 5$s$ bands, which is extremely sensitive to the local Sn-Se polyhedron environments \cite{Delaire_15SnSeAnharmonic_NatPhy}. Indeed, we have observed that VB along $\Gamma-Y$, which is contributed by Se $4p_{y}$-orbitals with negligible weight of Sn 5$s$ states, shows high degree of consistency between the ARPES results and DFT calculations (Figure \ref{Fig2}e). For the modelling of the pudding-mold VB, it is also important to take into account the unique point dislocation defects, which have drastically changed the 2D nature of ideal SnSe as evident by the quantum oscillation results. In the vicinity of these point defects, the local Sn-Se bonding network are likely to be disturbed by the reversing of stacking sequences between neighbouring MLs, leading to changes in the hybridisation of the Se $4p$- and Sn $5s$-orbitals.

Our results provide insights into the self-hole doping mechanism in SnSe, which is controlled by randomly distributed SnSe$_{2}$ micro domains inside the bulk. An alternative strategy to introduce self hole doping in SnSe without the formation of intercalating SnSe$_{2}$ is vacancy engineering. As shown in Table \ref{table:growth}, by using non-stoichiometric growth flux with insufficient Se, SnSe single crystals (SF12) with RT $\rho$ and hole concentration comparable to stoichiometric SF11 have been prepared, which suggests strong $p$-doping effect by Se vacancies. A systematic transport study on Se vacancy doped $p$-SnSe is now under way. The highly anisotropic quantum localization phenomena in $p$-SnSe also remind us that beyond the extraordinary thermoelectric performance, SnSe offers rich physics to be explored, which are rooted in the binary BP-type puckering structure.

\section{Methods}
\textbf{Single crystal synthesis.} Twelve batches SnSe single crystals were synthesized. The starting materials are Sn and Se elements form Alfa Aesar (99.999\%), which are are stored in a glove box filled with argon. For growth, they are mixed together according to the mole ratio $1:1$ or $1:0.95$ for different batches (details shown in Supplementary Table S1). The mixture was then sealed in a evacuated quartz tube with argon pressure below $10^{-2}Pa$. Three techniques are employed to grow SnSe single crystals, $i.e.$, self flux, Bridgeman, and physical vapor deposition (PVD), in a tubular furnace and a two-temperature zone furnace, respectively. As shown in Figure S1, for the self flux method, the growth flux is placed in the highest $T$ zone in the furnace; for the Bridgeman method, the tube end holding the flux is placed in the low $T$ zone, while the other end is placed in the high temperature zone to form a $T$ gradient for the growth section; for the PVD method, we synthesized SnSe crystal by fast-cooling self flux method first. The grown crystals with $\sim 1\%$ SnSe$_{2}$ were ground thoroughly and then sealed again. The tube was placed in the two-temperature zone furnace with a $T$-gradient of $50K$. The detailed growth curves for different batches of samples are summarized in Table S1.

\textbf{Angle-resolved photoemission spectroscopy measurements.} ARPES measurements were performed at the ``Dreamline'' beamline of the Shanghai Synchrotron Radiation Facility equipped with a Scienta D80 analyser. The energy resolution was set to 10~meV and the angular resolution was set to $0.2^\circ $. The samples were cleaved $in~situ$ and then measured at 13 K in a vacuum better than $5 \times 10^{-11}$ Torr. The ARPES data were collected using linearly horizontal-polarized lights with a vertical analyser slit.

\textbf{X-ray diffraction.} The crystallographic orientations and lattice structure of SnSe single crystals were determined by X-ray Bragg diffractions using a PAN-alytical X'Pert MRD diffractometer equipped with Cu $K_{\alpha}$ radiation and a graphite monochromator.

\textbf{Temperature-dependent transport measurements.} Different batches of SnSe single crystals were cut into rectangular shapes with the armchair and zigzag edges. Typical sample dimensions are $2\times1\times0.2 \mathrm{mm}^3$, corresponding to the $b$-,$c$-, and $a$-axes respectively. Due to the $p$-type doping, silver paint or epoxy are not suitable to make electric contacts. Instead, we spot welded $25\,\mathrm{\mu m}$-diameter Pt wires to our samples using a single pulse Sunstone welder to prepare the standard six-contact Hall-bar configurations. For metallic samples, the contact-resistance is less than $10\Omega$. All the transport properties were measured in a Oxford-14T cryostat, using Keithley 2400 and 2182A as current sourcemeter and nanovoltagemeters, respectively. Seebeck coefficients were measured with a temperature gradient of $<0.5K$, using two $E$-type miniature thermocouples to measure the $\Delta T$.

\textbf{Density-functional theory calculations.} The electronic structure of SnSe was calculated with the Vienna \textit{ab initio} simulation package (VASP) \cite{VASP_Kresse_PRB93,VASP_Kresse_PRB96} by the method of the projector augmented wave\cite{DFT_Blochl_PRB94} and the generalised gradient approximation (GGA) \cite{GGA_Perdew_PRL96}, to take the exchange-correlation potential as well as spin-orbit coupling into account. The plane-wave cutoff energy is set at about 400 eV and the k-point sampling is performed by the Monkhorst-Pack scheme \cite{MPscheme_Monkhorst_PRB76}. The total energy is ensured to be converged within 0.002 eV per unit cell.

\textbf{Non-contact atomic force microscopy.} We used ambient non-contact AFM (nc-AFM) on a multi-mode Park NX10 system to characterize the surface morphology of freshly cleaved SnSe single crystals. The point dislocation intensity is analysed by AFM imaging of 20 random locations spread over the whole crystal surface. For q-plus AFM, we conducted in-situ cleavage of bulk SnSe crystal followed by heating up the sample at 490 K for 2 hours to create clean surface required for high-resolution nc-AFM measurement. The turning-fork based q-plus sensor was employed to perform the nc-AFM experiment at 4.3K in ultrahigh vacuum conditions ($< 2\times 10^{-10}$ Torr). Q-plus nc-AFM imaging of SnSe crystal was measured in the constant amplitude mode, using the frequency shift as the feedback set point ($\Delta f$= -26 Hz, $f_{0}$ = 24.58 kHz, Q = 12,000, and Amplitude = 0.5 nm).

%Scanning tunneling microscopy and q-Plus based atomic force microscopy are measured on
\begin{acknowledgments} This work is supported by the National Science Foundation of China (Grant Nos. 11574264 and 11574337) and the National Key R\&D Program of the MOST of China (Grant No. 2016YFA0300204 and SQ2017YFJC010032-02). Y.Z. acknowledges the start funding support from the Thousand Talents Plan.
\end{acknowledgments}

\section*{Author contributions}
Z.X.S and Z.W. synthesised and characterised the single crystals. Z.W., Z.X.S, F.S., and Q.F.H. performed the transport measurements. C.C.F., Z.X.S, and D.W.S. carried out the ARPES experiments. C.Q.H and Y.H.L. did the DFT calculations. Z.Z.Q., Q.F.H., H.Y.F. and J. L. conducted the nc-AFM experiments. Z.W., C.C.F., D.W.S and Y.Z. analyzed the experimental data and wrote the paper with input from all authors. Y.Z. and D.W.S. supervised the project.

\bibliography{SnSe}

\end{document}